# Neurotoxicity of Silver Nanoparticles and Non-Linear Development of Adaptive Homeostasis with Age

**Anna A. Antsiferova** [1,2,*,‡], **Marina Yu. Kopaeva** [1, ‡], **Vyacheslav N. Kochkin**[1], **Alexander A. Reshetnikov**[1], **Pavel K. Kashkarov**[1,2,3]

[1] National Research Center "Kurchatov Institute", 1, Akademika Kurchatova sq., 123182 Moscow, Russia;
[2] Moscow Institute of Physics and Technologies, 9, Institutskii Lane, Moscow Region, 141700 Dolgoprudny, Russia
[3] Department of Physics, Lomonosov Moscow State University, 1, GSP-1, Leninskiye Gory, 119991 Moscow, Russia
* Correspondence: antsiferova_aa@nrcki.ru
† These authors contributed equally to this work.

**Abstract:** For the first time in the world behavioral functions of laboratory mammals exposed to silver nanoparticles were studied with the regard to age. Silver nanoparticles coated with polyvinylpyrrolidone with the size of 8.7 nm were used in the present research as a potential xcenobiotic. Elder mice adapted to the xcenobiotic better than younger animals. Younger animals demonstrated more drastic anxiety than the elder ones. Thus, it is concluded that adaptive homeostasis non-linearly changes with the age. Presumably, it may improve during the prime of life and start to decline just after certain stage.

## 1. Introduction

Adaptive homeostasis by definition is a transient expansion or contraction of the homeostatic range for any given physiological parameter in response to exposure to subtoxic, non-damaging, signaling molecules or events, or the removal or cessation of such molecules or events [1]. In the other words adaptive homeostasis is a capability of an organism to adapt to a damaging factor, for instance, to toxic compounds. It is assumed that adaptive homeostasis decreases in direct ratio to age incensement [2]. However, this is a rather questionable point. The other researchers believe that blossom of vitality falls on the first 1/3 of the maximal lifespan, i.e. middle age, which is followed by gradual decline of physiological functions [3]. Therefore, presumably, adaptive homeostasis may also be improved until about the first 1/3 of the lifespan and then be faded out (Fig. 1). Importantly, Fig. 1 demonstrates the total lifespan not as average but as a maximum life expectancy. Thus, the 'prime of life' point corresponds to 42-45 years old for a human.

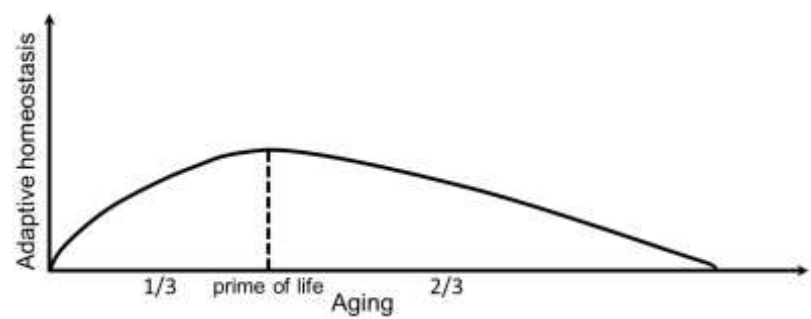

**Figure 1.** Adaptive homeostasis development with the age during the whole maximal lifespan.

Nowadays silver nanoparticles (AgNPs) are one of the most sought-after commercial nanotechnology products in the market [4]. They are widely used in medicine, pharmaceutics, food, cosmetic and light industries due to their excellent antiseptic properties [5-7]. At the same time, AgNPs may cause toxicity in healthy tissues as well. AgNPs sized smaller or larger than 10 nm could both cause neuronal cell death after entering the brain. It was shown that inflammation and increased oxidative stress followed by apoptosis are likely to be the main mechanisms of AgNPs toxicity [8,9]. Silver nanoparticle's toxicity was observed in testes after six months of oral exposure of Sprague Dawley rats [10]. Also, silver nanoparticles had cytotoxic effects on HepG2 cell line and primary liver cells of mice in *in vitro* study [11]. There is a large number of studies confirming silver nanoparticle's toxicity in regard to different cell types and tissues in *in vitro* and *in vivo* experiments. There is no doubt that silver nanoparticles may be considered as a xenobiotic for human and animals in certain dosages.

AgNPs demonstrate special bioaccumulative properties. They can accumulate in different tissues of an organism, particularly, in the brain [12-15] and disturb cognitive and behavioral functions [16-19] such as violations in the formation and consolidation of memory, learning, a decrease of social behavior and locomotor activity, anxiety increase. Thus, we showed before that behavioral functions of laboratory mice exposed to silver nanoparticles demonstrate 3-staged changes such as anxiety increase at the first, development of adaptation mechanism, which was detected by the increase of exploration behaviour, and finally disturbance of long-term contextual memory [16]. It is clear that widespread use of silver nanoparticles may be not safe enough and even harmful for human and the environment.

Toxic action of silver nanoparticles towards living organisms with the regard to age is not studied well. However, there is a sole work where dietary exposure of Drosophila to 12.5 nm AgNPs during early life with the regard to age as well as AgNPs mechanisms of toxicity are investigated [20]. It was shown that the exposure to AgNPs trigger multiple adverse effects on functional aging, including shortened lifespan, age-dependent decline, and loss of stress resistance and intestinal integrity. AgNPs inactivate antioxidant pathways in old, not young animals, increasing susceptibility to ROS in old age. Mammalian lifespan is longer and more complex than that in insects. Therefore, it is necessary to conduct such studies with model mammals to extrapolate the data to humans.

Thus, we investigated influence of silver nanoparticles onto behavioral functions of mammals at different ages as well their accumulation in different tissues. This work's general objective was to determine the direction of the adaptive homeostasis vector change.

## 2. Materials and methods

A food supplement Argovit S (Vector-Vita, Novosibirsk, Russia) recommended for GUT disease treatment [21] was used as AgNPs. The nanoparticles were coated with polyvinylpyrrolidone. Polivinylpyrrolidone is a hydrophilic polymer that forms hydrogen bounds with water molecules and provides solubility and stability of the nanoparticles in water media. The size and stability of the AgNPs were studied by Dynamical Light Scattering (DLS) (Malvern Zetasizer Nano ZS, Malvern, Great Britain). The shape of the AgNPs was studied by transmission electron microscopy (TEM) (Thermo Fisher Scientific, Waltham, MA, USA).

Male mice C57BL/6 obtained from "Stolbovaya" branch of the Federal Medical Biological Agency of Russia were used as a mammalian model. Behavioral functions of the animals were studied in the Open Field (OF), Elevated Plus Maze (EPM), and Light-Dark Box (LDB). Silver content in animal's organs was measured by Instrumental Neutron Activation Analysis (INAA) with the use of research nuclear reactor IR-8 (Moscow, Russia) with a capacity of 8 MWth and a gamma spectrometer (ORTEC, Oak Ridge, TN, USA).

## 3. Experiment scheme

### 3.1. Study of nanoparticle's morphology

The optimal concentration of AgNPs solution was selected to prevent multiparticle scattering and to ensure a sufficient counting time. The original 10 mg/ml solution was diluted 10-, 50- and 100-fold with deionized water (Milli-Q, USA). The optimal concentration was established as 0,1 mg/ml on the base of the empirical research. This concentration was further used to study AgNPs size. 10 ml of the initial solution was poured into a sealed container, which was kept in the dark at +2 °C during 1 year to study the nanoparticle's stability. The concentration of 0,02 mg/ml was applied for AgNPs visualization by TEM.

### 3.2. Animal treatment and preparation administration

The mice were kept in individual cages during the whole experiment with unlimited access to food and water in rooms with automatically maintained temperature of 23 ± 2 °C and a 12/12-h day/night cycle. The room's humidity was controlled at 45 ± 10%. The animal's body mass was controlled weekly.

There were two groups of mice 3 months apart in age. The mice of the first group were introduced into the experiment since the age of 2 months (younger, n=20). The mice of the second group were introduced into the experiment from the age of 5 months (elder, n=24). Both groups were divided into control (control 1, n =10; control 2, n=12) and experimental subgroups (younger, n=10; elder, n=12).

AgNPs were introduced daily orally with drinking water in the amount of 50 μg per day per animal during 60 days. The drinkers were weighted weekly to control consumed amount of liquid. It was noticed that the animals consumed equal amount of liquid (3.8 mL per day) at the constant air humidity. Based on this, the required amount of silver nanoparticles was dissolved in the pure water Osmoteck 40-3-2 (OOO "Pharmsystemy", Besedy, Moscow region, Russia).

The mice were about 5 (younger) and 8 (elder) months old by the end of the experiment and the data obtaining time points. Conditionally assuming a maximum mouse lifespan over 36 months (no less than 36 months) [22], it can be easily calculated that the prime of life point (Fig. 1) is 12 months. Thus, according to the second theory considered in the Introduction an improving of adaptive homeostasis should take place between regarded ages. According to the first theory, a gradual decline of the adaptive homeostasis would be observed in the regarded period of mice life.

### 3.3. Behavioral tests

Behavioral tests for locomotor activity, exploration behavior and anxiety assessment started since 61th day of the experiment. At the same time, we continued the experimental animal exposure to AgNPs in order to exclude possible elimination processes. Videotracking equipment and software EthoVision XT 8.5 (Noldus Information Technology, Wageningen, The Netherlands) was used for detecting the animal's position and movement.

#### 3.3.1. Open Field

Locomotor activity was measured in the OF on the 61st day of the experiment similarly to [16]. Briefly, a mouse was placed in the middle of the plastic circular arena (d = 120 cm, h = 45 cm) and allowed to explore OF for 5 min. The mouse's behavior was recorded via an overhead Sony video camera (Tokyo, Japan). For analysis, the OF was divided into three virtual areas – central (d = 60 cm), peripheral (r = 10 cm, near the wall) and intermediate (between central and peripheral). The following parameters were compared: distance traveled (total, central, intermediate and peripheral), latent period of exit from the central area, time spent in the areas, central and intermediate areas entries, number of rearings.

### 3.3.2. Elevated Plus Maze

The EPM is used to measure anxiety-like behavior in rodents [23]. The test is based on the natural tendencies of rodents to avoid open or elevated places counterbalanced with their innate curiosity to explore areas that are new to them. The EPM consists of four elevated arms (two open arms, 30 × 5 × 0.5 cm; two closed arms, 30 × 5 × 15 cm) which radiate from a central platform (5 × 5 cm), forming a plus shape. Testing was carried out on the 63rd day of the experiment similarly to [16]. Briefly, mice were placed in the central platform, facing an open arm and were allowed to explore the apparatus for 5 min. The mouse's behavior was recorded via an overhead Sony video camera (Tokyo, Japan). The following parameters were compared: total distance traveled, latent period of the first approach in the closed and the open arms, open and closed arm duration, open and closed arms entries, head dipping, number of rearings.

### 3.3.3. Light-Dark Box

LDB is conventional test for assessment of anxiety-like behavior in the laboratory rodents, based on the approach–avoidance conflict [24,25]. The LDB (50 × 50 × 40 cm) was divided into two parts: 1/2 was painted black, covered by the lid and separated from the white compartment by the wall containing an opening (8 × 6 cm) at the floor level. Testing was carried out on the 65th day of the experiment similarly to [16]. Briefly, animals were placed individually in the middle of the light part and were allowed to explore the apparatus for 10 min. The mouse's behavior was recorded via an overhead Sony video camera (Tokyo, Japan). Mice were compared by the following parameters: the latency to cross to the dark chamber, the time spent in the light chamber, average speed and distance traveled in the light chamber, number of peeking out from the dark chamber, number of transitions between compartments.

### 3.4. Silver content measurement in organs and brain departments

After conducting a battery of behavioral tests, the animals (n = 6 – 8 in each soubroup) were anesthetized with isoflurane (Baxter, Baxter Healthcare Corporation of Puerto Rico, USA) and decapitated; the organs (liver, spleen, lungs, kidneys, brain, heart, testes) and whole blood were collected. This protocol was based on other widely used procedures [26, 27]. Tissue samples were dried in a drying chamber RedLine RF 53 (Binder GmbH, Germany) for 72 h at 75°C for further irradiation in the channel of a nuclear reactor. Dried samples were placed in hermetically sealed, and numbered with a moisture-resistant marker polyethylene containers (Eppendorf, Hamburg, Germany) at volumes of 0.2, 0.5 and 2 mL. At this time, reference samples were prepared for simultaneous irradiation in the channel of a nuclear reactor and for measurements by the method of comparison with a standard sample [28]. Cotton wool was placed in the same plastic containers (to maintain the identity of the geometry factor), and a known amount of the standard sample of silver (100 or 1000 ng per sample) (LenReaktiv, Saint Petersburg, Russia) was added for this. The containers were left open, air dried for 48 h and then hermetically sealed. Compact reference samples were also prepared. For this, a known amount of state standard sample of silver was placed on paper disks and air-dried. Next, plastic containers and reference sam-ples were placed into aluminum cases. Each aluminum case contained one reference sample with the same geometry factor as the experimental samples, as well as 84 compact samples. Aluminum cases were suspended in a vertical channel of the nuclear reactor and irradiated for 24 h at a neutron flux of $10^{12}$cm$^{-2}$s$^{-1}$. After irradiation, the cases were kept in biosecurity for the decay of highly active short-lived isotopes, and then gamma-spectrometric studies of the samples were carried out for the activities of the radioactive isotope $^{110m}$Ag with a half-life of 250 days.

### 3.5. Statistical analysis

Data are expressed as mean ± SEM. GraphPad Prizm 6.01 software (La Jolla, San Diego, CA, USA) was used for statistical analysis, with statistical significance at P < 0.05. The nonparametric Kruskal–Wallis ANOVA with a *post hoc* Dunn's test or Mann–Whitney U test were employed.

## 4. Results

### 4.1. Nanoparticles

According to DLS data the mean size of AgNPs was 8,7±0,4 nm (Fig. 2). The nanoparticles were stable because their mean size did not change after storage for one year. TEM demonstrated that the shape of the nanoparticles was quasi spherical (Fig. 3).

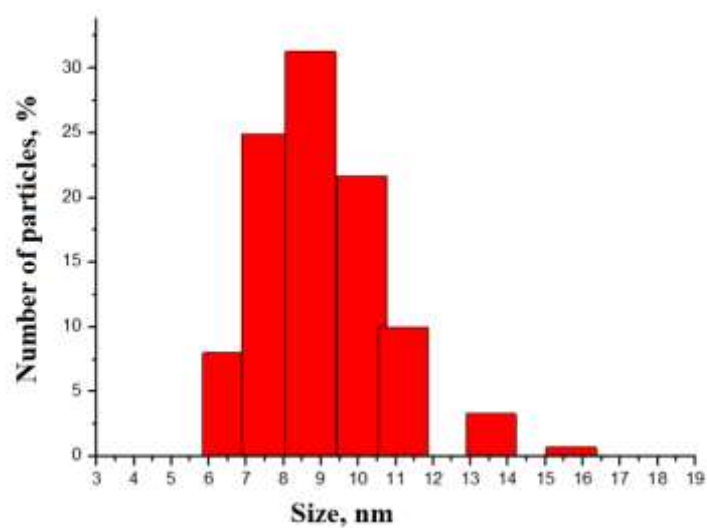

**Figure 2.** Distribution of the number of Agovit-S particles by size according to the DLS data. Mean size is 8,7±0,4 nm.

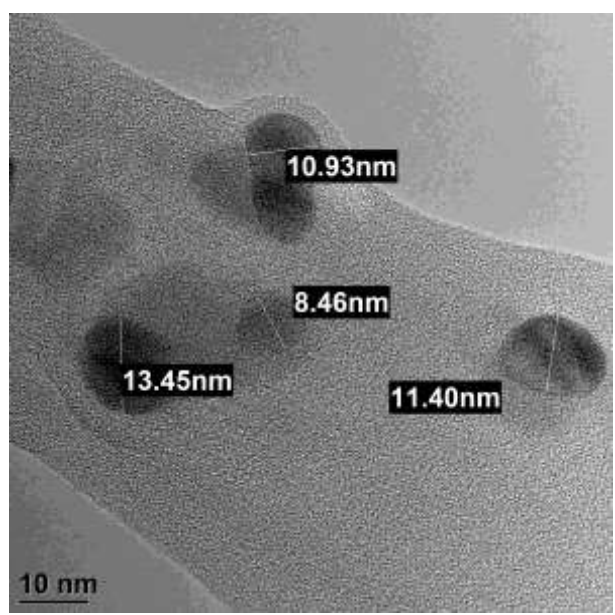

**Figure 3.** TEM image of Argovit-S AgNPs. The shape of nanoparticles is quasi spherical.

*4.2. Physiological characteristics*

    All the mice were gradually growing. No statistically significant difference in their body weight between experimental and control groups was detected (Fig. 4).

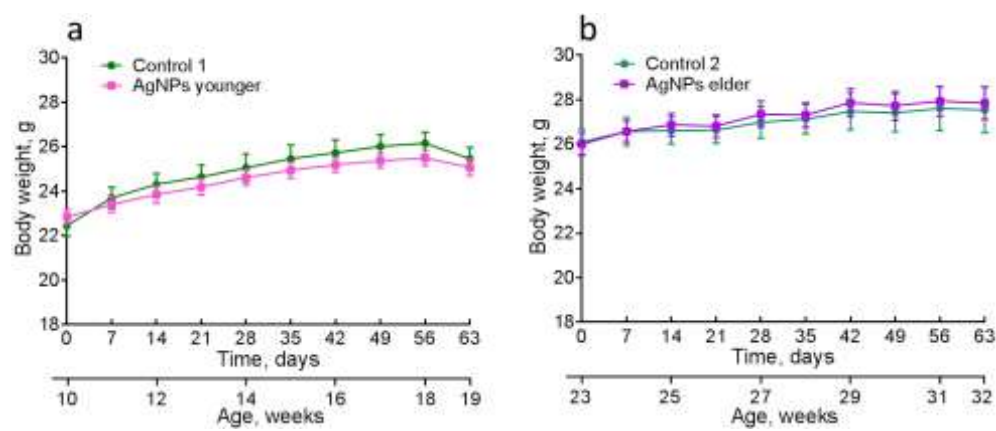

**Figure 4.** Changes in the body weight of mice with the age: a – younger vs control 1, b – elder vs control 2.

*4.3. Behavioral functions of animals*

    Statistically significantly different results of behavioral tests for all groups are present below (Fig. 5-7)

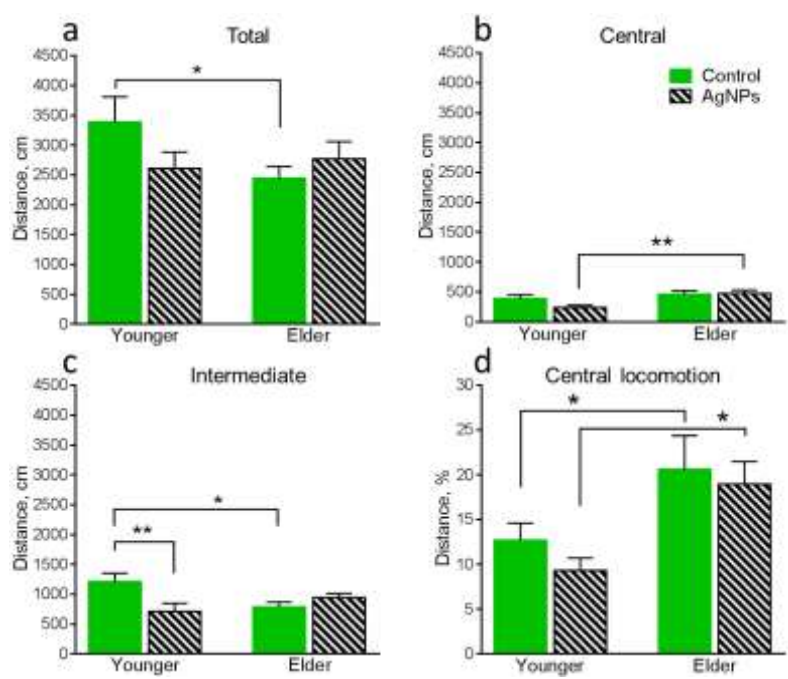

**Figure 5.** Assessment of behavioral functions in the OF: a - total distance moved in the OF, b – distance moved in the central area of the OF, c – distance moved in the intermediate area of the OF, d - central locomotion in the OF. Values are presented as mean ± SEM. *p < 0.05, **p < 0.01.

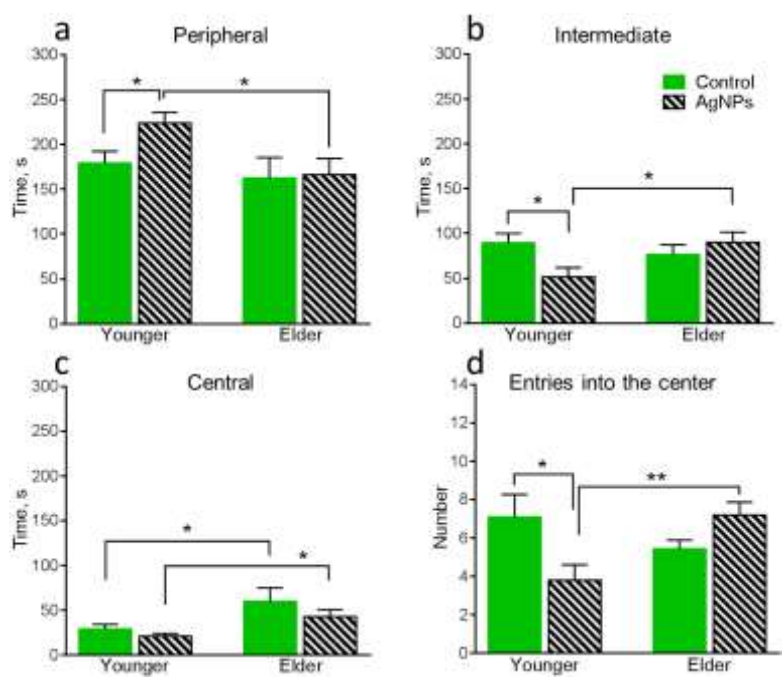

**Figure 6.** Assessment of behavioral functions in the OF: a - time spent in the peripheral area of the OF, b – time spent in the intermediate area of the OF, c – time spent in the central area of the OF, d – number of entries into central area of the OF. Values are presented as mean ± SEM. *p < 0.05, **p < 0.01.

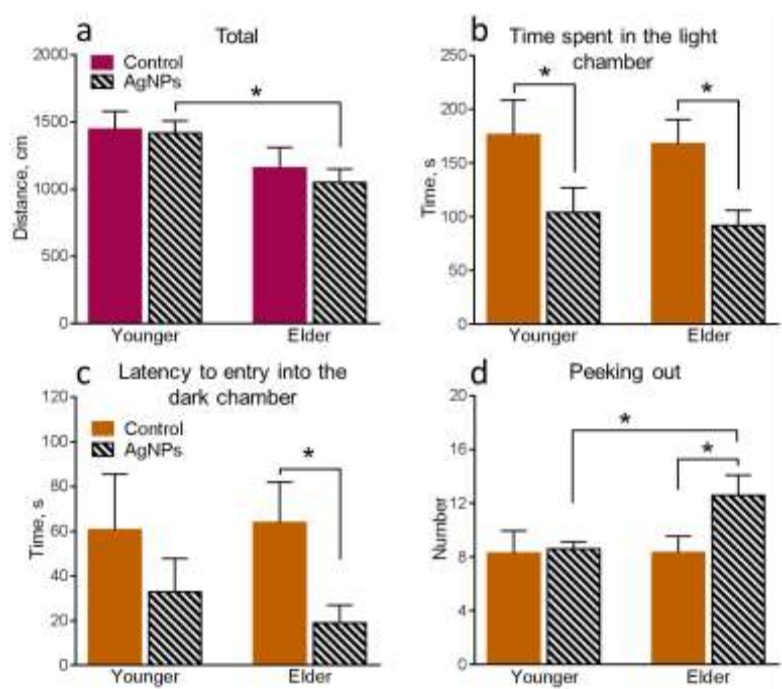

Figure 7. Assessment of behavioral functions in the EPM and LDB: a - total distance moved in the EPM, b – time spent in the light chamber of the LDB, c – latency to entry into the dark chamber of the LDB, d – number of peeking out from the dark chamber of the LDB. Values are presented as mean ± SEM. *p < 0.05.

Firstly, it is necessary to compare behavioral functions of control groups in order to detect the influence of age-related changes in the absence of exogenous perturbation in the form of AgNPs. The locomotor activity decreased with the age that can be noticed from reduction of the total distance moved (Fig. 5a) and distance moved in the intermediate area (Fig. 5b) of the OF. Also, anxiety decreased simultaneously, which can be seen from increase of the time spent in the central area of the OF field (Fig. 6c).

Further, we determined the influence of oral AgNPs exposure to young mice in comparison with the control 1. The xenobiotic exposure significantly increase anxiety in mice that can be noticed by several parameters simultaneously. The distance moved in the intermediate area of the OF (Fig. 5c), the time spent in this area of the OF (Fig. 6b) and the number of entries into center of the OF (Fig. 5d) decreased, while the time spent in the peripheral area of the OF (Fig. 6a) increased. Also, time spent in the light chamber of LDB decreased (Fig. 7b).

Behavioral functions of the elder mice exposed to AgNPs were compared with control 2. It can be seen that the mice exposed to AgNPs remain more anxious than the control ones. It can be noticed by decrease of the time spent in the light chamber (Fig. 7b) and the latency to entry into the dark chamber (Fig. 7c) of the LDB. However, the number of anxiety indicators decreased in total. Also, increase of peeking out from the dark chamber of LDB (Fig. 6d) demonstrated increase of exploration behavior which is a positive characteristic.

The results of comparative analysis of behavior of younger and elder mice both exposed to AgNPs point to decrease of locomotor activity with the age, which can be seen from the decrease of the total distance moved in the EPM (Fig. 7a). Decrease of the locomotor activity is typical for elder animals without the exogenic perturbation as was shown above. General decrease of anxiety in elder group can be noticed in increase of the distance moved in central (Fig. 5b) and intermediate (Fig. 5c) areas of the OF, decrease of the time spent in the peripheral area of the OF (Fig. 6a), as well as increase of the number of entries into the center of the OF (Fig. 6d) Elder group of mice also demonstrated increase of exploration behavior which was seen in the higher number of peeking out from the dark chamber of LDB in elder animals compared to young ones (Fig. 7d). As shown above, anxiety decrease is also typical for elder control animals. Increased locomotor activity and somehow higher anxiety of younger mice can be explained by keeping them in individual cages, which is unnatural condition for mice living in flocks in the wild nature [29]. Keeping animals in the individual cages 3 months longer could neutralize the negative factor and lead to their adaptation to such conditions.

4.4. AgNPs accumulation in the internal organs

The content level of silver in all control groups was lower than the detection limit. It is due to that silver is not an essential element for mammalian organisms. The content of silver in some heart, spleen and kidney samples were lower than detection limit, which was an obstacle for the statistical analysis. Therefor we did not consider those organs further. Figs 8 and 9 show comparative accumulation of silver in the internal organs and blood of the experimental animals with the regard to age.

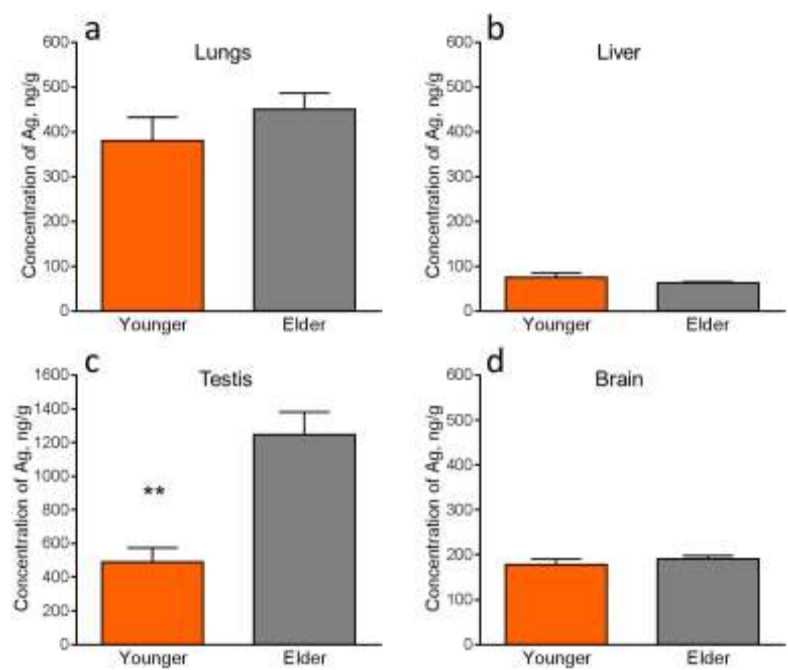

Figure 8. Accumulation of silver in the internal organs of younger and elder mice exposed to silver nanoparticles. Values are presented as mean ± SEM. **p < 0.01.

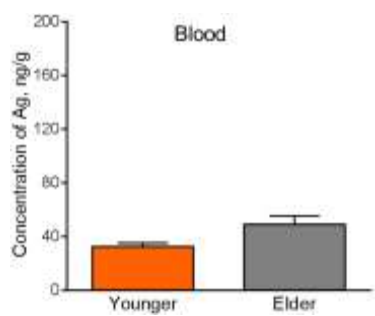

**Figure 9.** Content of silver in the blood of younger and elder mice exposed to silver nanoparticles.

Figure 9 demonstrates that the level of silver accumulation in testes was statistically significantly higher in elder than in younger mice. It can evidence about physiological changes based on the slowing down the toxin elimination processes with age.

Concentrations of silver in blood for both ages were lower than in the other considered organs. Low content of silver in blood is typical for subacute, subchronic and chronic exposures to AgNPs [15] and corresponds to static equilibrium. High content of silver in blood is observed at acute exposure to AgNPs [30].

## 5. Discussion

In general, all the elder mice in comparison to younger ones demonstrated certain improvement of behavioral functions. Hereby, the xenobiotic in the form of AgNPs could show a hormetic effect related to improvement of adaptive homeostasis in elder animals [2]. Probably it correlates with the concept that improvement of physiological functions and adaptive homeostasis takes place until 1/3 of the maximal lifespan. Herewith, it is likely that physiological and cognitive functions are related in direct proportion. It was shown earlier that physiological and cognitive functions are associated [31], i.e., mental health and physiology are inextricably linked. Good physical characteristics correlate with good cognitive abilities [32]. Also, it is known that cognitive functions improve till the middle age that is confirmed by higher IQ levels of 35-40 years old humans in comparison to other age groups [33]. Cognitive processes have been supposed to be among the mechanisms that drive behavior [34] and it is likely that good cognitive functions correlate with good behavioral functions and vice versa. A correlation between cognition and behavior was found in epilepsy [35]. Thus, mental and physiological functions are the facets of the same crystal and a good quality of adaptive homeostasis is straightly proportional to good physiology and cognition.

We observed neurotoxicity decrease of AgNPs with ageing manifested in the improvement of certain behavioral functions of elder animals in comparison to younger ones. This indicated the increase of adaptive homeostasis quality with the age in the period of vitality blossom. Herewith, we observed more significant silver accumulation in hippocampus and testes of elder animals compared to that in younger ones. Presumably, higher silver accumulation in those organs may cause different toxic effects, which were not considered in the present work. Bioaccumulation of the potential xenobiotic may cause toxic effects later, after the beginning of decline of adaptive homeostasis and weakening of its function as well. Thus, AgNPs accumulation in hippocampus may be hazardous for memory [19, 36]. We observed decline of long-term contextual memory in mice after 180 days of exposure to AgNPs of the same trade mark [16,19]. It should be carefully studied in the future research.

Accumulation of AgNPs in testes may cause various toxic effects as well [10, 37-39]. Exposure to AgNPs may influence mice fertility [40]. Also, AgNPs may be transported from the exposed during pregnancy females to the offspring via placental and mammary gland barriers [41]. Thus, transited AgNPs, silver ions and granules may accumulate in the offspring's organisms and cause toxic effect in it. Besides direct toxic effects on the tissues where AgNPs accumulate they can exert toxicity at the whole organism lever, which can be explained within the framework of neurovisceral integration concept [42]. Within the concept of neuravisceral integration toxicity in different organs may lead to certain hemisphere's function disturbance [43] and dysfunctions in the brain may lead to broad range of dysfunctions in the organism. Thus, AgNPs may be potentially harmful for the offspring.

## 6. Conclusions

There are two contradictory concepts about adaptive homeostasis development with the age. The first of it states that it declines proportionally to age. The second concept affirms that adaptive homeostasis non-linearly changes during life span. It increases with the age during vitality blossom and starts to decline since 1/3 of the maximal lifespan.

In the present work we observed the change of behavioral functions of laboratory mice as a mammalian model under the influence of a xenobiotic in the form of AgNPs with the regard to age. Elder animals demonstrated improvement of behavioral functions in comparison to younger ones despite increase of the accumulated silver testes and not in the brain of them. This phenomenon can be explained by the better adaptation of the animals to the potential xenobiotic in elder age. Adaptive homeostasis non-linearly change with the age. Thus, our work confirms the second concept.

**Author Contributions:** Conceptualization, A.A.A. and M.Yu.K.; methodology, M.Yu.K; DLS, A.A.A.; biological experiment, M.Yu.K., instrumental neutron activation analysis, V.N.K. and R.A.A.; formal analysis, M.Yu.K., R.A.A.; analysis and interpretation, A.A.A.; writing, A.A.A; editing, M.Yu.K; supervision, P.K.K; project administration, A.A.A. and K.V.N.; funding acquisition P.K.K. All authors have read and agreed to the published version of the manuscript.

**Funding:** The study was partially financed by the Russian Foundation for Basic Research (grant No. 21-315-70016) and partially supported by the Ministry of Science and Higher Education of the Russian Federation, the contract 075-15-2021-709, unique identifier of the project RF-2296.61321X0037 (equipment maintenance).

**Institutional Review Board Statement:** The study was conducted according to the rules of the Ministry of Health of the Russian Federation (№ 267 of 19.06.2013), and approved by the Local Ethics Committee for Biomedical Research of the National Research Center "Kurchatov Institute" (No 01 from 10.02.2017).

**Informed Consent Statement:** Not applicable.